\documentstyle[pre,aps,epsfig,multicol]{revtex} 
 \begin{document} 
 \draft                                                     
 \title{Disappearance of Spurious States in Analog Associative Memories} 

 \author{Yasser Roudi and Alessandro Treves} 
   
 \address{SISSA - Programme in Neuroscience, via Beirut 4, 34014 Trieste, Italy, 
 {\tt yasser@sissa.it} } 
 \date{\today} 
 \maketitle 
   
 \begin{abstract} 
 We show that symmetric $n$-mixture states, when they exist, are almost never 
 stable in autoassociative networks with threshold-linear units. Only with 
 a binary coding scheme we could find a limited region of 
 the parameter space in which either 2-mixtures or 3-mixtures are stable 
 attractors of the dynamics.                                                                 
 \end{abstract} 
 \begin{multicols}{2}[]
\section{Introduction}
 Autoassociative networks are useful models of one of the basic operations 
 of cortical networks\cite{Rol+98}. `Hebbian' plasticity on recurrent 
 connections, e.g. in the higher level areas of sensory cortex and in the 
 hippocampus, is the crucial ingredient for autoassociation to work, 
 with real neurons \cite{Tre+91}. Neural network models, although very 
 simplified and abstract, allow a comprehensive analysis, indicating whether 
 associative memory retrieval can proceed safely, or whether it must face 
 dynamical hurdles, such as `spurious' local minima in a free-energy landscape. 
The dynamics of such networks, in the simplest models, is governed by a number 
$p$ of dynamical attractors, each of which corresponds to a distribution
of neural activity, i.e. a pattern, which represents a long term 
memory. Memory is stored by superimposed synaptic weight changes,
and the basic operation proceeds by supplying the network with an
external signal that acts as a cue, correlated, perhaps only weakly, with 
a pattern, and which leads through attractor dynamics to the retrieval
of the full pattern.\\

 How smoothly can such an operation proceed, and how wide
are the basins of attraction of the $p$ memory states? Clearly, these
issues depend critically on whether other attractors exist, that could
hinder or obstruct retrieval  As a crude example, if the cue is 
 correlated with the image of a mule, the net may be able to retrieve either 
 a horse or a donkey, if no "mixed" attractor exist. If instead the 
 encoding procedure has, unintentionally, created a spurious 
 attractor for the mule itself, the network will likely be stuck 
 in such a mixed memory state. In a slightly more complicated model 
 endowed with some topographic mapping of visual space, a horse cue 
 and a donkey cue might be presented simultaneously in neighbouring 
 positions. If they are too close in visual space and spurious 
 attractors exist, this topographic map might retrieve two mules next 
 to each other. Returning to nets without spatial structure and considering 
 for simplicity only symmetric mixtures of patterns embedded with equal strengths,
 there are obviously $p(p-1)/2$ 2-mixtures, $p(p-1)(p-2)/6$ 3-mixtures,
 and so on. Do they correspond to stable attractors, and as such do they
 influence the network dynamics?\\
 
In addition, connectionist modelers have proposed to 
 describe in terms of spurious states certain psychiatric dysfunctions 
 \cite{Ami+89}. Speech disorders in schizophrenic patients, for instance, 
 might arise from the existence of a large number of spurious states, 
 that obstruct the retrieval of correct patterns \cite{Hoff+87}. 

 In their seminal investigation of the Hopfield model \cite{Hop82}, 
 Amit, Gutfreund and Sompolinsky found that while symmetric mixtures of an 
 even number of patterns are unstable, odd mixtures and the spin glass phase 
can be stable, in a certain region of phase space \cite{Ami+85}. 
In the Hopfield model, though, neurons are modelled as binary units, 
 and correspondingly each distribution of activity, in particular 
 each memory pattern, is a binary vector. Either or both of these aspects might 
 be essential in producing the additional minima in the free-energy 
 landscape. Real neurons behave very differently from binary units 
 in many respects, a basic one being that their spiking activity, once filtered 
 with a short time-kernel \cite{note1}, is better approximated by an analog variable. 
 Threshold-linear units reproduce this graded nature of neural response, 
 yet still allow for a simple and complete statistical mechanics analysis 
 of autoassociative network models \cite{Tre90}. With threshold-linear units, 
 the memory patterns encoded in the synaptic weights can still be taken to 
 be binary vectors, but can also be taken to be drawn from a distribution   
 with several discrete activity values, or from a continuous distribution \cite{Tre+91}. 
 Exponential distributions, in particular, can be argued to be not far 
 from experimentally observed spike count distributions \cite{Lev+95}. 

 The question of mixture states in analog nets was first 
 addressed in \cite{Wau}, arguing that the multiple local minima 
 of the spin glass phase are fewer in number in an associative net of units with 
 more continuous (sigmoid) transfer function. Later it was found, 
 considering threshold-linear units, that are both realistic and amenable to analytical treatment
\cite{Tre+91}, that the region of stability of the spin glass 
 phase is severely restricted with such units \cite{Tre91b}, again indicative of a 
 general smoothing of the free-energy landscape with analog variables. Although these 
 analyses provide a good starting point, they are not complete in the sense that they 
 did not show what will happen to $n$-mixture states with $n$ small (the ones 
 relevant to models of schizophrenia), and what is the 
 effect of different coding schemes, that is pattern distributions. Here, we consider 
 instead symmetric $n$-mixtures, with $n=2,3,\dots$, and we consider non-binary memory 
 vectors. Also, from the biological point of view, it is important to study nets 
 with diluted (incomplete) connectivity, which are much more realistic descriptions 
 of cortical \cite{Bra+91} and hippocampal networks \cite{Tre+92}, where the probabiliy 
 of a recurrent connection between any two units may be of the order of a few percent. 

 In this manuscript we show that symmetric mixture states give rise to 
 dynamical attractors only in very restricted circumstances, in associative networks 
 of threshold-linear units, both with full and diluted connectivity. We have 
 analysed the validity of this statement in different coding schemes, and did not 
 find any stable mixture state at all, when memory patterns are not binary. 
 Essentially, we conclude that this type of spurious states are a pathological 
 feature of the simplified binary models considered in the initial studies. 
\section{threshold linear model}
 We use a model very similar to that analysed in \cite{Tre90}. We consider a 
 fully connected network of $N$ units, taken to model excitatory neurons. The 
 level of activity of unit $i$ is a dynamical variable $V_i\ge 0$, which corresponds 
 to the short time averaged firing rate of the neuron. Units are connected 
 to each other through symmetric weights. The specific covariance 'Hebbian' learning rule we consider 
 prescribes that the synaptic weight between units $i$ and $j$ be given as 
 \begin{equation} 
 J_{ij}=\frac{1}{Na^2}\sum_{\mu=1}^p\left(\eta_i^{\mu}-a\right)\left(\eta_i^{\mu}-a\right), 
 \end{equation} where $\eta_i^{\mu}$ represents the activity of unit $i$   
 in pattern ${\mu}$. Each $\eta_i^{\mu}$ is taken to be a quenched variable, drawn 
 independently from a distribution $p(\eta)$, with the constraints $\eta \ge 0$, 
 $\langle \eta \rangle_{\eta} = \langle \eta^2 \rangle_{\eta}= a$. As in one of the 
 first extensions of the Hopfield model \cite{Tso88}, we thus allow for the mean activity 
 $a$ of the patterns to differ from the value $a=1/2$ of the original model \cite{Tre90}. 

 The model further assumes that the input to unit $i$ takes the form 
 \begin{equation} 
 h_i=\sum_{j\neq i}J_{ij}^cV_i+\sum_{\nu}s^{\nu}\frac{\left(\eta_i^{\nu}-a\right)}{a}+ 
 b\left(\frac{1}{N}\sum_jV_j\right), 
 \end{equation} 
 where the first term enables the memories encoded in the weights to determine 
 the dynamics; the second term allows for external signals $s^{\nu}$ to cue the 
 retrieval of one or several patterns; and the third term is unrelated to the 
 memory patterns, but is designed to regulate the activity of the newtork, 
 so that at any moment in time $\frac{1}{N}\sum_i V_i = \frac{1}{N}\sum_i V_i^2 = a$. The activity 
 of each unit is determined by its input through a threshold-linear function 
 \begin{equation} 
 V_i=g(h_i-T_{thr})\Theta(h_i-T_{thr}) 
 \end{equation} 
 where $T_{thr}$ is a threshold below which the input elicits no output, $g$ is 
 a gain parameter, and $\Theta(...)$ the Heaviside step function. Units are 
 updated, for example, sequentially in random order, possibly subject to fast 
 noise. The exact details of the updating rule and of the noise are not specified 
 further, here, because they do not affect the steady states of the dynamics, 
 and we take the noise level $T$ to be vanishingly small, $T\to 0$. Discussions about 
 the biological plausibility of this model for networks of pyramidal cells can be 
 found in \cite{Tre+91,Ami91}, and will not be repeated here. 

 Subject to the above dynamics, the network evolves towards one of a set of attractor 
 states. In a given attractor the network may still wander among a variety of 
 configurations, but it reaches a stationary probability distribution of being 
 in any particular configuration. The average of any quantity over such `annealed' 
 probability distribution is denoted by $\langle \rangle$ (whereas $\langle 
 \rangle_{\eta}$ denotes the average over the quenched distribution $p(\eta)$ ). 
 To analyse such a model one can introduce, as in \cite{Tre90} the order parameters:
 \begin{eqnarray}
 x&=&\sum_{i=1}^{N}v_i\\   
 x^{\sigma}&=&\frac{1}{Na}\sum_{i=1}^{N}\eta_i^{\sigma}v_i-x\\    
 y_0&=&\frac{1}{N}\sum_{i=1}^{N}\langle V_i^2 \rangle \qquad\\
 y_1&=&\frac{1}{N}\sum_{i=1}^{N}\langle V_i \rangle^2. 
 \end{eqnarray} 
 where $x$ is simply the mean activity of the network, and  $x^{\sigma}$,the subtracted, or specific, 
 overlap of the current state of the network with each of the stored patterns.
 Two further parameters, \begin{eqnarray}
\psi&=& (y_0-y_1)T_0/T \\
 \rho&=&{py_1\over [N(1-\psi)^2]},  \end{eqnarray} 
can be defined as a function of $y_0$ and $y_1$, and play 
 a particularly useful role in the analysis in the limit we consider, $T\to 0$, 
 when one configuration dominates the annealed average, and $y_1\simeq y_0 + O(T)$. 
 The characteristic noise scale of the system is $T_0\equiv (1-a)/a$ \cite{Tre90}, 
 and we define the storage load $\alpha\equiv p/N$. In the limit $N\to 0, T\to 0$, 
 the system is thus characterized by the parameters $a$ (mean pattern activity, which 
 also parametrizes the coding sparseness \cite{Tre90} in the sense that decreasing 
 $a$ makes the code sparser), $\alpha $ (storage load), $g$ (gain) and $T_{thr}$ 
 (threshold). 
\section{Mean field solutions and their stability}
 We calculate the free energy using the replica trick, for 
 symmetric $n$-mixture states (where $n$ overlaps take the same non-zero 
 value, and the rest are zero) elicited by external signals $s^1=\dots=s^n=s$. 
 These signals can be purely transient, so that at steady state $s=0$, but we 
 consider a non-zero steady value for the sake of generality. We look for symmetric 
 states, characterized by non-zero $\hat{x}^1=\dots=\hat{x}^n=\hat{x}$. 
 The saddle point equations reduce to \cite{Tre90}:
\begin{eqnarray}
  x&=&g'\ll \int_{h>T_thr}Dz(h-T_{thr})\gg\\
  x^{\sigma}&=&g'\ll(\frac{{\eta}^{\sigma}}{a}-1)\int_{h>T_thr}Dz(h-T_{thr})\gg\\   
  \psi&=&T_0 g'\ll \int_{h>T_thr}Dz\gg\\
  y_0&=&(g')^2 \ll \int_{h>T_thr}Dz(h-T_{thr})^2\gg\\
  {\rho}^2&=&\frac{\alpha y_0}{(1-\psi)^2}
  ,h_2=\frac{\alpha T_0}{2(1-\psi)}
  ,g'=\frac{g}{1-2gh_2}
\end{eqnarray} 
where now the input to each unit can be expresed as:
\begin{equation}
h=b(x)-\sum_{\sigma}x^{\sigma}+\sum_{\sigma}\frac{{\eta}^{\sigma}}{a}(x^{\sigma}+s^{\sigma})-zT_0\rho
\end{equation}
and the free energy reads:
\begin{eqnarray*}
f=&-&\frac{g'}{2}\ll
\int_{h>T_thr}Dz(h-T_{thr})^2\gg+\frac{1}{2}\sum_{\sigma}(x^{\sigma})^2\\
&+& xb(x)-B(x)+\frac{T_0}{2}{\psi}{\rho}^2
 \end{eqnarray*} 
If one defines new parameters 
$v=(\hat{x}+s)/(T_0\rho)$ (the specific 
 signal-to-noise ratio) and $w=[b(x)-n\hat{x}-T_{thr}]/(T_0\rho)$ (a sort of 
 uniform field-to-noise ratio), it is easy to show that the mean field equations can be reduced to :
 \begin{eqnarray} 
 E_1(w,v)&=&(A_1+\delta A_2)^2-\alpha A_3=0\label{E1}\\ 
 E_2(w,v)&=&(A_1+\delta A_2)(\frac{1}{gT_0(1+\delta)}-A_2)-\alpha A_2=0 
 \label{E2} 
 \end{eqnarray} 
 where 
 \begin{eqnarray} 
 A_1(w,v)&=&A_2(w,v) - \langle\int^+Dz\rangle_{\eta}\\ 
 A_2(w,v)&=&\frac{1}{nvT_0}\langle(\frac{\Gamma}{a}-n)\int^+Dz(w+v 
 \frac{\Gamma}{a}-z)\rangle_{\eta}\\ 
 A_3(w,v)&=&\langle\int^+Dz(w+v\frac{\Gamma}{a}-z)^2\rangle_{\eta}, 
 \end{eqnarray} 
 with $\Gamma = \sum_{\sigma=1}^n{\eta}^{\sigma}$and $\delta = s/\hat{x}$. 
 In the equations above, $Dz=\frac{dz}{\surd 2\pi}e^{-z^2/2}$ and the subscript $+$ 
 indicates that the $z$-average has to be carried out only in the range where 
 $w+v\frac{\Gamma}{a}-z>0$. In the following we take $\delta = 0$. Thus symmetric 
 $n$-mixture attractors exist if we can find stable solutions of Eqs.\ref{E1}, 
 \ref{E2}.
 
 To analyze the stability of the extrema of the free energy, one has to 
 study the hessian matrix 
 \begin{equation} 
 H_{\mu\nu}={\delta}_{\mu\nu}-\langle(\frac{{\eta}^{\mu}}{a}-1) 
 (\frac{{\eta}^{\nu}}{a}-1){\int^+Dz}\rangle_{\eta} 
 \end{equation} around the saddle point. 

 In general, for $n$-mixture states, there are three types of eigenvalues: 

 1. a non-degenerate eigenvalue, which decides the stability against a 
 uniform increase in the amplitude of the $n$ patterns that contribute 
 to the thermodynamic state (i.e. the 'condensed' patterns), while 
 the other overlaps remain zero. It is (for $\mu\neq \nu$)   
 \begin{equation} 
 {\lambda}_1=1-\langle(({\frac{{\eta}^{\mu}}{a}-1)}^2+n(\frac{{\eta}^{\mu}}{a}-1) 
 (\frac{{\eta}^{\nu}}{a}-1))\int^+Dz\rangle_{\eta} 
 \end{equation} 

 2. an eigenvalue of degeneracy $n-1$, associated with any direction which 
 tends to change the relative amplitude of the non-zero overlaps. It is (again 
 for $\mu\neq \nu$) 
 \begin{equation}                               
 {\lambda}_2=1-\langle(({\frac{{\eta}^{\mu}}{a}-1)}^2-(\frac{{\eta}^{\mu}}{a}-1) 
 (\frac{{\eta}^{\nu}}{a}-1))\int^+Dz\rangle_{\eta} 
 \end{equation} 

 3. the third eigenvalue, with degeneracy $p-n$, measures the stability 
 against the appearance of additional overlaps.   
 \begin{equation} 
 {\lambda}_3=1-T_0\langle\int^+Dz\rangle_{\eta}. 
 \end{equation} 

 \section{Different coding schemes} 
 In order to proceed further, we restrict the analysis to a number of specific 
 coding schemes, i.e., to different choices for the distribution $p(\eta)$. 
 We consider 
 \begin{eqnarray} 
 p(\eta)&=&a\delta({\eta}-1)+(1-a)\delta({\eta}), \quad {\rm binary}\nonumber\\ 
 p(\eta)&=&\frac{a}{3}\delta({\eta}-\frac{3}{2})+a\delta({\eta}- 
 \frac{1}{2})+(1-\frac{4a}{3})\delta(\eta), \quad {\rm ternary}\nonumber\\ 
 p(\eta)&=&4a{e}^{-2\eta}+(1-2a)\delta(\eta), \quad {\rm exponential}. 
 \end{eqnarray} 

 For small values of the 
 load $\alpha$ (and hence of the quenched noise $\rho$), Eq.\ref{E2} describes an 
 hyperbole, whose center depends on the value of $g$. Eq.\ref{E1} instead, for 
 small values of $\alpha$, is a closed curve in the quartant $w<0, v>0$, 
 so that with an appropriate choice of $g$ the two curves intersect at two points. 
 As $\alpha$ grows, the region $E_1(v,w)>0$ shrinks in size, until at a certain 
 value of $\alpha$, which depends only on $a,n$ and the 
 coding scheme, it reduces to a point and then disappears.

 We have investigated not just the existence but also the stability of solutions 
 for symmetric 2- and 3-mixture states.
 The solutions behave exactly in 
 the same manner in these two cases: for small values of $a$ both intersections 
 discussed above are unstable, in the sense that both $\lambda_1$ and $\lambda_2$ 
 are negative. This finding is confirmed by computer simulation, in which one of 
 the overlaps tends to grow, reaching the corresponding attractor, whereas the 
 other one (or the other two in the case of 3-mixtures) tend to zero. 
 Increasing the value of the sparsity parameter, one finds different results   
 with binary coding and with other types of coding. 

 \begin{figure}[h] 
 \epsfxsize 9cm 
 \centerline {\epsfbox{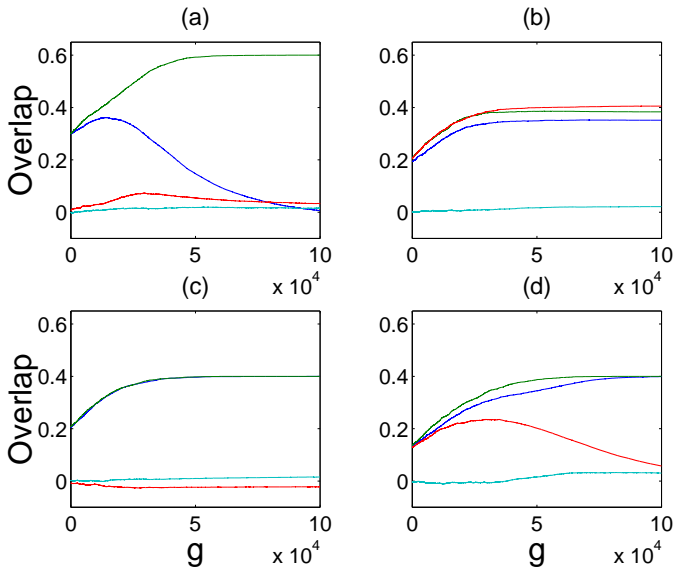}} 
 \caption{Computer simulation result $N=10000, p=5$ and 
 (a,b): $g=1.2, a=0.4$; (c,d): $g=3, a=0.6$. In (a,c) the initial state 
 was correlated equally with 2 patterns, in (b,d) with 3.} 
\label{fig1}
 \end{figure} 

 Let us consider binary coding first. 
 After a range of $a$-values with only one unstable eigenvalues (${\lambda}_1$ 
 or ${\lambda}_2$), one finds a range where genuinely 
 stable solutions can be found. Thus the retrieval of mixture patterns is possible 
 for binary coding, as can be seen in the simulations shown in Fig.\ref{fig1}. 

\begin{figure}[h] 
 \epsfxsize 6cm 
 \centerline {\epsfbox{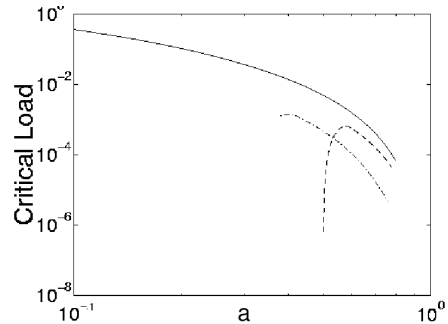}} 
 \caption{Storage capacity as a function of sparseness for the single pattern 
 states (full line), compared to the region of existence and stability of 
 2-mixtures (dashed line) and 3-mixtures (dashed dotted line), for the binary 
 coding scheme, in a fully connected network with threshold-linear units. Points 
 denote the $a, \alpha$ values used in the simulations of Figure 1.}
\label{fig2} 
 \end{figure}

 The exact stability region in the $(\alpha, a)$ plane differs for 2-mixtures and 
 3-mixtures. In both cases, it is delimited to the right by the `critical load' 
 $\alpha_c(a,n)$, i.e. the value at which the island with $E_1(v,w)>0$ shrinks to zero, 
 and to the left by the load $\alpha $ beyond which no intersection with both 
 $\lambda_1>0$ and $\lambda_2>0$ can be found. Fig.\ref{fig2} illustrates these stability 
 regions, compared with the critical load for the pure attractor states, 
 as in \cite{Tre90}. 

\begin{figure}[h] 
 \centerline{\hbox{\epsfig{figure=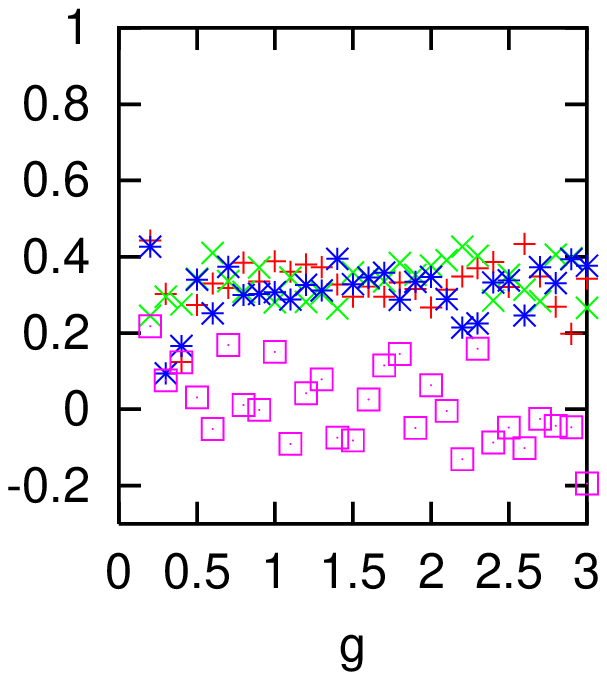,width=4cm,angle=270}\epsfig{figure=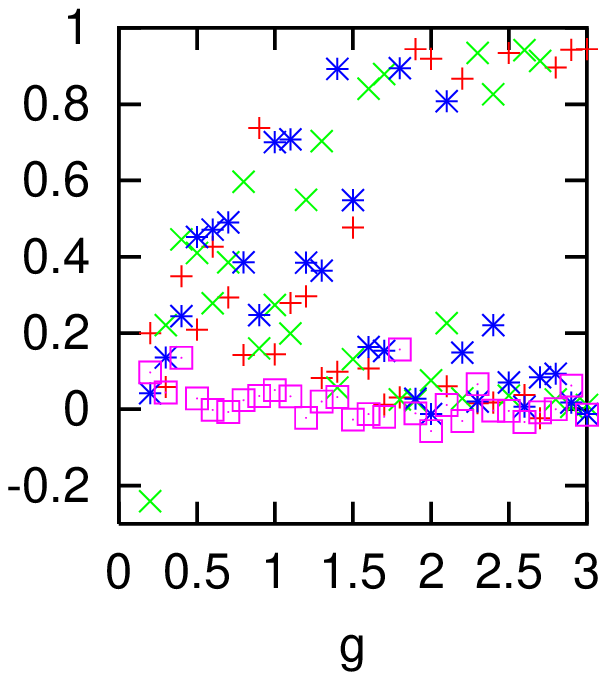,width=4cm,angle=270}}} 
 \caption{The final overlaps with each of 4 stored patterns, averaged over the last 
 400 updates of the whole network, for left: binary and right: ternary coding, in both 
 cases with N=10000, a=0.5, p=4. 3 patterns have initially a non-zero overlap 
 with the activity of the network and retain it, for most $g$ values, in the binary 
 coding case, while a single pattern is always selected in the ternary case.}
 \label{fig3}
 \end{figure}

 For ternary and exponential coding, the solutions of the saddle point equations 
 remain unstable even for very high values of the sparsity parameter $a$. 
 Again, this was verified by computer simulations. Fig.\ref{fig3} illustrates the different 
 situation occurring with ternary and binary coding, by considering a 
 very low load and a sparsity value for which stable solutions for 3-mixtures are 
 easily found in the binary case. Note that the `critical load' for 3-mixtures 
 would be considerably higher with ternary patterns (not shown); the fact is 
 that at each position of the intersection, either ${\lambda}_1$ or ${\lambda}_2$ 
 or both turn out to be negative. 
 This complex behaviour of eigenvalues will be discussed elsewhere in more detail. 

\section{diluted case}
 We have also extended the analysis to a highly diluted network \cite{Der+86}. 
 In this case the number of patterns that can be stored scales with the
number $C$ of connections each unit receives, rather than with the number of units $N$.
One then redefines the load parameter as $\alpha \equiv p/C$.
The essential difference introduced by the sparse (i.e. diluted) connectivity
is that noise has less of an opportunity to reverberate along closed loops.
In fact the signal, which during retrieval is simply contributed by the `condensed'
patterns, propagates coherently and proportionally to $C$, independently of 
the density of feedback loops in the network. The fluctuations in the overlaps with
the undecondensed patterns, which as $T\to 0$ represent the sole source of
noise, propagate coherently along feedback loops, giving rise to the
amplifying factor $1/(1-\psi)$ of the fully connected case.
For a given load (fixed $\alpha$), diluted connectivity reduces therefore
the influence of this `static' noise, and performance is better than in the
fully connected case with $N-1=C$. In particular with the extreme dilution, that
is, if the condition $\frac{c}{Ln(N)} \rightarrow 0$ is satisfied, one can
neglect correlations among the $C$ inputs to a given unit \cite{Der+86}, and
the mean field equations become \cite{Tre91a}: 
 \begin{eqnarray} 
 E_1(w,v)&=&(A_2+\delta A_2)^2-\alpha A_3=0\\ 
 E_2(w,v)&=&(\frac{1}{gT_0(1+\delta)}-A_2)=0. 
 \end{eqnarray} 

 Examining again the stability matrix, we find that the mixture solutions, that were 
 present with binary coding and large values of $a$, still survive. By the token, 
 the results for ternary and exponential coding 
 are not affected, in the sense that no stable solutions can be found even in the 
 highly diluted case. 
\section{conclusion}
 The conclusion is that the existence of stable mixture states in a restricted 
 region of the parameter space  should be regarded as almost a pathological 
 feature, resulting from binary coding. If one considers 
 mixture states as spurious states, to be avoided, then one notes that 
 the introduction of analog variables, a more realistic description of neural 
 activity, goes a long way towards disposing of spurious states, just 
 as it almost eliminated the spin glass phase \cite{Tre91b}. The remaining 
 region of stability of spurious states is definitely eliminated by 
 non-binary coding schemes, that further contribute to smooth the free-energy 
 landscape. This result casts doubts upon e.g. models of schizophrenia that 
 are based on the existence of spurious attractors.   

 These results may well have implications in domains outside computational 
 neuroscience. The smoothness of the free-energy landscape is a crucial 
 features of many interacting systems used to map optimization problems, 
 such as the travelling salesman\cite{Hop+86} or the graph matching problem \cite{Fu+86}. 
 Optimization generally fails if the dynamics gets stuck into local minima. 
 Our result indicates that undesired local minima may be eliminated 
 by a combination of analog variables and coding schemes, which may in some cases be manipulated while 
 mapping the problem at hand onto a dynamical system.

 \end{multicols} 

\begin{references} 
 \bibitem{Rol+98} 
 E.T. Rolls, A. Treves, {\em Neural Networks and Brain Function}, (Oxford U.P., Oxford, 1998). 
 \bibitem{Tre+91} 
 A. Treves, E.T. Rolls, Network {\bf 2} 371 (91) 
 \bibitem{Ami+89} 
 D.J.Amit, {\em Modeling Brain Function}, (Cambridge U.P., Cambridge, 1989). 
 \bibitem{Hoff+87} 
 R. E. Hoffman, Arch. of Gen. Psych.  {\bf 44} 178 (1987) 
 \bibitem{Hop82} 
 J.J. Hopfield, Proc. Natl. Aca. Sci. USA {\bf 79} 2554 (82) 
 \bibitem{Ami+85} 
 D.J. Amit, H. Gutfreund, H. Sompolinsky, Phys. Rev. {\bf A 32} 1007 (85), 
 D.J. Amit, H. Gutfreund, H. Sompolinsky, Ann. Phys. (N.Y.) {\bf 173} 30 (87) 
 \bibitem{note1} 
 Such short time kernel may be thought to correspond to the conversion of 
 presynaptic spikes into postsynaptic potentials. 
 \bibitem{Tre90} 
 A. Treves, Phys. Rev A {\bf 42} 2418 (90) 
 \bibitem{Lev+95} 
 W.B. Levy, R.A. Baxter, Neural Comp. {\bf 8} 531 (96); 
 R.J. Baddeley {\em et al}, Proc. Roy. Soc. (London) {\bf B 264} 1775 (97); 
 A. Treves {\em et al}, Neural Comp. {\bf 11} 611 (99) 
 \bibitem{Wau} 
 F.R. Waugh {\em et al}, Phys. Rev. Lett {\bf 64} 1986 (90) 
 \bibitem{Tre91b} 
 A. Treves, J. Phys. A: Math. Gen. {\bf 24} 2645 (91) 
 \bibitem{Bra+91} 
 V. Braitenberg, A. Sch\"utz, {\em Statistics of the Cortex: Anatomy and Geometry}, 
 (Springer, Berlin, 1991) 
 \bibitem{Tre+92} 
 A. Treves, E.T. Rolls, Hippocampus {\bf 2} 199 (92) 
 \bibitem{Tso88} 
 M.V. Tsodyks, M.V. Feigel'man, Europhys. Lett. {\bf 6} 101 (88) 
 \bibitem{Ami91} 
 D.J. Amit, M.V. Tsodyks, Network {\bf 2} 259, 275 (91) 
 \bibitem{Der+86} 
 B. Derrida, E. Gardner, A. Zippelius, Europhys. Lett. {\bf 4} 167 (87) 
 \bibitem{Tre91a} 
 A. Treves, J. Phys. A: Math. Gen. {\bf 24} 327 (91) 
 \bibitem{Hop+86} 
 J.J. Hopfield, D.W. Tank, Science {\bf 233} 625 (86) 
 \bibitem{Fu+86} 
 Y. Fu, P.W. Anderson, J. Phys. A {\bf 19} 1605 (86) 
 \end{references}
 \end{document}